**Mahyar Taj Dini, V. Yu. Sokolov**

# INTERNET OF THINGS SECURITY PROBLEMS

The rapid development of "smart" devices leads to explosive growth of unprotected or partially protected home networks. These networks are easy prey for unauthorized access, the collection of personal information (including from surveillance cameras), interference in the operation of individual devices and the entire system as a whole. In addition, existing solutions for managing a smart house offer work in the cloud, which in turn reduces the availability of the system and simultaneously increases the risk of the unscrupulous use of personal information by the service provider (up to the sale of data to a third party). This article examines the existing access technologies, their weaknesses, and offers solutions to improve the overall security of the system with a local IoT gateway and virtual subnets.

**Introduction**

The Internet of Things (IoT) is essentially a construct where machines (cloud and data center-based apps) and common devices (such as watches, toasters, thermostats, body monitors and cars) are connected to each other via the public Internet. Within the IoT, common devices are controlled and monitored remotely using wireless networks for the most part, while data flows between the cloud and traditional data centers for analysis and manipulation.

While this may suffice as an appropriate technical definition, it is hardly appropriate in respect to how the consumer must understand the IoT and specifically how it will directly affect their personal privacy. In this regard, the worst-case scenario is one where the consumer forfeits all of their privacy due to ignorance or complacency, and then has every detail of their personal lives made available to anyone who wants to pay for this information from the app provider or one of the many data brokers who will dominate the secondary market for IoT data. Many of these perpetrators will then target these same consumers with specific adverts and offers, as well as performing behavioral experimentation, usually without the consumer's knowledge, much less specific consent. For further insights on these types of scenarios, see my previous privacy corner postings on data brokers and social experimentation.

The potential for such ubiquity (billions to trillions of devices) of IoT seems like a foregone conclusion at this point. But there are multi-dimensional privacy challenges which must be surmounted if this truly is going to become a reality. To get ahead of these challenges the privacy engineering community (via National Institute of Standards and Technology) is currently involved in intense discussions as to how to "engineer in" the right privacy regime, which will provide users (consumers) with direct control over a wide range of their own personal privacy settings as well as creating auditing and measuring schemes to ensure compliance with both user settings as well as regulatory mandates.

Privacy engineering is a very real challenge, and there are multiple paths in the IoT where a privacy regime must be monitored and maintained:
- The device (data generator, data receiver and aggregation point).
- The Internet (multi-directional data transport).
- The cloud (data manipulation and aggregation point).
- The machine (application services, big data repositories, analytics and more).

Each path requires appropriate privacy protections to be engineered into it, with user control wherever appropriate (device, machine and others) while being maintained along its entire length (virtual and physical). High levels of encryption, redundancy and security will be necessitated to counter threats in flight as well as at the endpoints. There will also be regulatory controls and adherence monitoring, which must be facilitated along these same pathways. Most of these will fall under the auspices of FTC (US), Data Privacy Act (EU), and other regulatory bodies and statutes across the world.

In parallel with the need for comprehensive privacy, security and compliance capabilities, the IoT is entirely predicated on new business models, which disrupt conventional solutions. An enabler of this disruption is the cost model component, which dictates low inherent costs in the devices, and all other components of the value chain. These cost models will not be conducive to "out of band" controls via bolt on solutions. Engineering-in privacy as part of the device and other pathway structures will be the only path to success in which cost efficiencies are maintained while compliance is assured along the way.

IoT extends the "Zone of Privacy Vulnerability" for consumers to the innermost reaches of their lives. They will be monitored (and potentially manipulated) every second of every day, no matter where they are (unless completely off the grid) or what they are doing (awake or asleep). No longer is there a buffer zone in the form of endpoints such as PC's, tablets or mobile phones. In the world of IoT, devices are attached to the consumer and embedded in everything around them, streaming data (and secrets) continuously to a variety of benign and potentially nefarious recipients and third parties. It is paramount that we take this into account as we develop a privacy strategy for IoT [1].

**Default Passwords**

All smart devices are built based on some operating system, sometimes these operating systems are based on Unix, Linux or sometimes they are so simple as they can just run some very specific actions by the way all operating systems has different user privileges and security levels, and this can be one of security potential in this way that manufactures who built these devices or who developed or customized these operating systems can set Default passwords, but now we should know what is the risk potential exactly ,in this article we mention two risk side of such situation.

Users mostly don't change default passwords and it can be very dangerous even in not targeted attack but in random attacks also the y can be victim of hackers and easily they can lost their privacy and security.

And second, users don't know about these different passwords, because mostly manufactures use some of these passwords as their backdoor or just for their service centers, by the way without telling clients such habits are prohibited by international privacy laws like article 8 of the European Convention on Human Rights, which was drafted and adopted by the Council of Europe in 1950 and meanwhile covers the whole European continent except for Belarus and Kosovo, protects the right to respect for private life: "Everyone has the right to respect for his private and family life, his home and his correspondence". Through the huge case-law of the European Court of Human Rights in Strasbourg, privacy has been defined and its protection has been established as a positive right of everyone. Data privacy laws are converging in the EU, helped by the National data protection authorities and the Data Protection Directive adopted in 1995. Article 17 of the International Covenant on Civil and Political Rights of the United Nations of 1966 also protects privacy: "No one shall be subjected to arbitrary or unlawful interference with his privacy, family, home or correspondence, nor to unlawful attacks on his honor and reputation. Everyone has the right to the protection of the law against such interference or attacks."

And same instance on different regions and countries, but still such hidden passwords levels are available for example, A backdoor has been found in devices made by a Chinese tech firm specializing in VoIP product. Or According to the leaked information, FortiOS operating system, deployed on Fortinet's FortiGate firewall networking equipment, includes an SSH backdoor that can be used to access its firewall equipment. Anyone with "Fortimanager_Access" username and a hashed version of the "FGTAbc11*xy+Qqz27" password string, which is hard coded into the firewall, can login into Fortinet's FortiGate firewall networking equipment. This issue affected all FortiOS versions from 4.3.0 to 4.3.16 and 5.0.0 to 5.0.7, which cover FortiOS builds from between November 2012 and July 2014, and of course many different other instances.

**Limited Permission Access**

Mostly companies don't take full permission to clients, Some manufacturers or carriers may try to refuse you warranty service if you find a specific privilege access level to manage their devices. For example, on Mobile Devices we see such issue, iPhone are Jailed, Android devices are not Rooted and then clients cannot access to file systems and make changes in deep of system files or even install some application to give information about their devices. Then without security analyzing we cannot approve these manufactures don't use clients data's or even they don't have any backdoor on such devices. specially smart devices are involved to our life 24/7 and even homes TV or even refrigerators and are on such risks.

**Privacy for Sale**

Some companies also collect and analyze information about users' "tweets, posts, comments, likes, shares, and recommendations." While many of these details were already available on the data companies' websites, the lawmakers used the letters as a chance to raise awareness about an industry that they said has largely "operated in the shadows."

And now with growing IoT devices such companies can work easier because they have ear & eye everywhere, in our homes, in industry or even in very sensitive structures, today IoT devices are working everywhere and collecting data for manufactured companies are easy, and because of reasons that we wrote before even it is hard to detection. In this phase, we can see to side of risk, first side is non personal data's, for example, the Czech Republic based security software vendor AVG Technologies recently updated its privacy policy. The objective of the changes, according to the company, was to explain in a more transparent manner to their users how it intends to use what it calls "non-personal information". The new privacy policy will take effect on 15 October 2015.The company defines "non-personal data" as data that cannot be linked to the identity of users in any way. The new privacy policy explains that the company might collect and sell this information to third parties) or many other samples.

Over the past years or so, a huge amount of attention has been paid to government snooping, and the bulk collection and storage of vast amounts of raw data in the name of national security. What most of you don't know, or are just beginning to realize, is that a much greater and more immediate threat to your privacy is coming from thousands of companies you've probably never heard of, in the name of commerce. They're called data brokers, and they are collecting, analyzing and packaging some of our most sensitive personal information and selling it as a commodity to each other, to advertisers, even the government, often without our direct knowledge. Much of this is the kind of harmless consumer marketing that's been going on for decades. And as we mentioned even it is possible for them to sell personal data's specially for spying or watching for example: released to the public on March 7, 2017, the first set of documents has been called "Year Zero" by Wikileaks. It is said to include details of the CIA's global hacking program, its malware, and zero-day exploits for a number of devices. And in these devices even smart TV's are available and the exploits are about remote voice sniffing or remote capturing.

**Data Manipulation**

Today, more than 92 percent of critical business records are generated, managed and stored electronically, creating efficiencies and cost-savings for businesses. Unfortunately, digital information can be easily deleted, altered and/or manipulated. For businesses, the burden of proof is on the company to ensure and attest to the accuracy and credibility of their electronic business records. This ability to prove the integrity of critical business records becomes especially important in litigation where executives are often called upon to support their claims of ownership of any discoverable records, as well as verify their history of creation and use. Electronic records have been proven to have been manipulated in cases ranging from stock options fraud to loan fraud to intellectual property disputes.

Some recent examples of actual cases surrounding the manipulation of electronic records include top executives at a successful technology company attempted to alter electronic records to hide a secret options-related slush fund to cover the tracks of their backdating options scheme.

A prominent real estate developer received an electronic version of a loan agreement to print and sign. Rather than just signing the document, he made subtle changes to it in order to make the terms of the loan more favorable to himself. The changes went undetected for a year until the loan was refinanced or in some other possible examples for example some companies can manipulate clients data's like their power-supply counters, or any frauds [2].

**Suggestion as Solutions**

As first solution, we are suggesting using clouds. The cloud is a very broad concept, and it covers just about every possible sort of online service, but when businesses refer to cloud procurement, there are usually three models of cloud service under consideration, Software as a Service (SaaS), Platform as a Service (PaaS), and Infrastructure as a Service (IaaS). The comparison of cloud models is shown on Fig. 1.

**Infrastructure as a Service**

IaaS provides the infrastructure such as virtual machines and other resources like virtual-machine disk image library, block and file-based storage, firewalls, load balancers, IP addresses, virtual local area networks etc. Infrastructure as service or IaaS is the basic layer in cloud computing model.

Common examples: DigitalOcean, Linode, Rackspace, Amazon Web Services (AWS), Cisco Metapod, Microsoft Azure, Google Compute Engine (GCE) are some popular examples of IaaS.

**Platform as a Service**

PaaS or platform as a service model provides you computing platforms, which typically includes operating system, programming language execution environment, database, web server. It is a layer on top of IaaS as the second thing you demand after Infrastructure is platform.

Common examples: AWS Elastic Beanstalk, Windows Azure, Heroku, Force.com, Google App Engine, Apache Stratos.

**Software as a Service**

In a SaaS you are provided access to application services installed at a server. You don't have to worry about installation, maintenance or coding of that software. You can access and operate the software with just your browser. You don't have to download or install any kind of setup or OS, the software is just available for you to access and operate. The software maintenance or setup or help will be provided by SaaS provider company and you will only have to pay for your usage.

Common examples: Google Apps, Microsoft office365, Google docs, Gmail, WHMCS billing software.

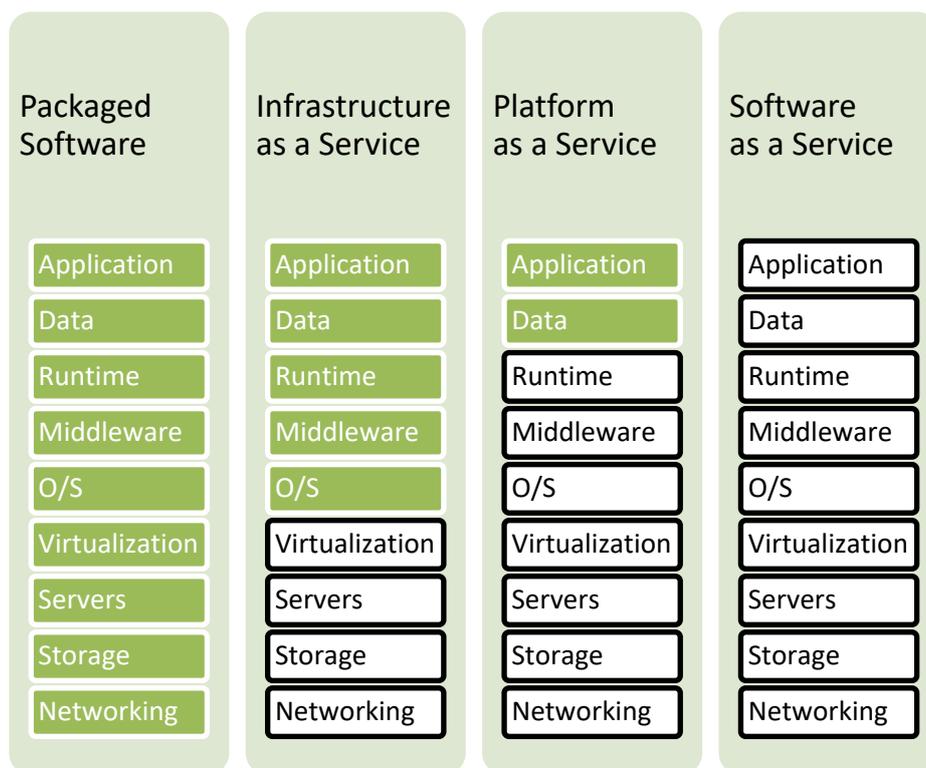

Fig. 1. Comparison of cloud models

In our solution data from IoT sensors to cloud should be encrypted, and using cloud as infrastructure to process data's without knowing which data they are receiving then even PaaS is not suitable in this case because in PaaS, operating system, middleware and runtimes are also depends on cloud then again risk of backdoor are still available and too much then we are offering to use IaaS (Fig. 2).

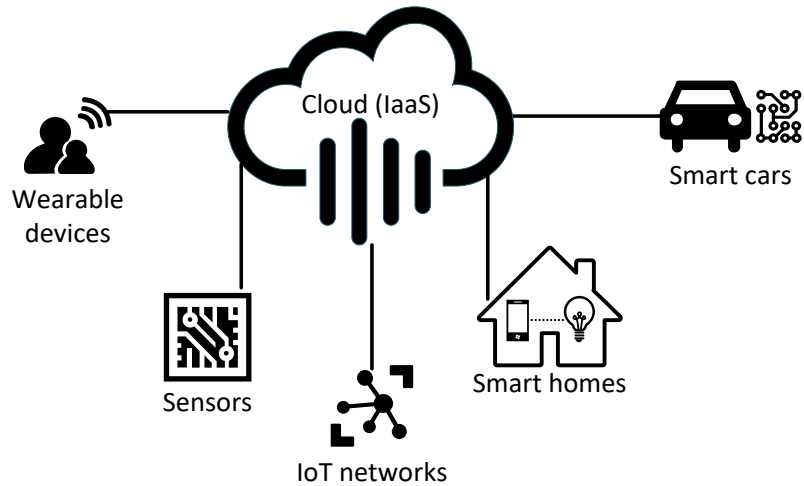

Fig. 2. Classical scheme of IoT system

Another solution is using IoT gateway and collecting data on it like shown in Fig. 3.

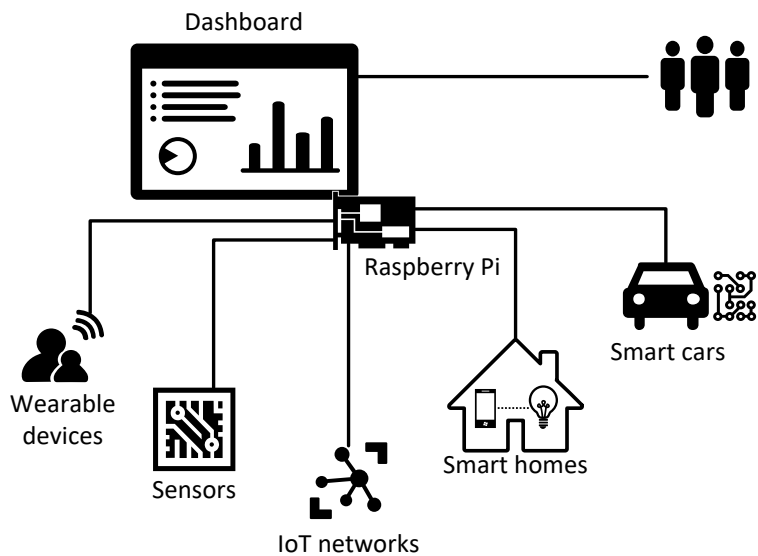

Fig. 3. "Cloud" based on the home system

As the abilities and needs of devices proliferate, it is often not possible to have them communicate directly with systems. Some sensors and controllers don't support energy-intensive protocols like Wi-Fi or Bluetooth. Some devices aggregate data so that it is overwhelming and invaluable in its raw form and they are all connecting to a variety of public and private networks.

An IoT gateway performs several critical functions from translating protocols to encrypting, processing, managing and filtering data. If you imagine an IoT ecosystem, a gateway sits between devices and sensors to communicate with the cloud.

IoT gateways help to bridge the gap between operations and IT infrastructure within a business. They do this by optimizing system performance through the operational data they gather and process in real-time in the field or at the network edge.

IoT gateways can perform a number of enhancements on the OT and IT silos:

- High scalability—they are able to take intelligent data from the datacenter or cloud and push into the field or network edge.
- Lowering costs—end-point devices needn't have as high processing power, memory or storage since the gateway does this all for them.
- Faster production—an accelerated and more advanced production line can decrease time-to-market significantly.
- Reduce telecommunications cost—less M2M communication means a smaller network and (WAN) traffic.
- Mitigate risks—gateways can isolate devices and sensors that aren't performing before they cause bigger problems for the production line.

**Adding a Layer of Security**

As the number of devices and sensors grow, so does the number of communications that will take place over a combination of public and private networks. Communications between the 'things', the gateway and the cloud therefore must be secure in order to prevent any data tampering or unrestricted access.

This will usually happen through a PKI infrastructure, whereby every "thing" that communicates is given an identity, that is, a pair of cryptographic keys (or Digital Certificate) which allows communication to be encrypted. This can be quite a handful to manage without the help of an IoT gateway.

Assuming you have a tool which manages all of your device certificates, you need the gateway to help mediate the on-boarding of devices (installation of certificates and provisioning of identity)

How to Secure an IoT Gateway

There are three key core principles of security—confidentiality, integrity and authentication. You will need to ensure that all communications between the gateway and devices are meeting each of the three principles while communication is happening in the internal and external networks.

It is also worth noting that the gateway is often the first to be attacked because of two reasons:

It has a higher processing power, which it can use to run more intensive applications. More power means better software, but better software means more vulnerabilities for a hacker to exploit.

Because of its location as an Edge device between the internet and the intranet, the gateway is the point of entry for any threat vector (as well as a system's first line of defense).

Our recommendations on securing an IoT gateway device involve three steps.

*Step 1. Identity for the Gateway Device*

The first step would be to give your gateway device an identity (by using an X.509 Digital Certificate). Any external entities connecting to the gateway will now be able to verify the identity of the gateway which is now enabling HTTPs or NTLS protocols. Commands being issued to devices or sensors in the field are now coming from a trusted device.

*Step 2. Enable "Strong" Identity for the Gateway Device*

Because your gateway device is vulnerable to physical tampering, private keys can be extracted and cloned leaving your gateway device vulnerable to spoofing or even man-in-the-middle (MITM) attacks.

In order to prevent this, you would have to use extra security measures, such as embedding a Trusted Platform Module (TPM) device into your Gateway, using a PUF (Physical Un-clonable Function). This would securely store the private keys of all Digital Certificates, making sure they never leave the gateway.

*Step 3. Use the Gateway to Provision Identity to Your Ecosystem*

Now that you have enabled strong identity in your gateway device, you need to think about having strong identity for the devices and sensors in the field. Because some of these are likely unable to connect to the internet, provisioning identity through a Certificate Management Service without a Gateway will be difficult [3].

Instead, we can use the gateway as a trusted security mechanism to secure anything that is connected to the gateway (on the intranet). The gateway acts as a proxy between the platform (CA

Services) and the devices in the field. As with the device itself you would expect this to happen using the standard PKI infrastructure, that is, an X.509 certificate through a private hierarchy.

Now the gateway and devices are secure and therefore all the communication in your intranet is secure. Therefore, you have security, confidentiality and authentication, allowing your IoT ecosystem to be end-to-end secured using a PKI infrastructure [4].

Another solution can be like diagram as shown in Fig. 4.

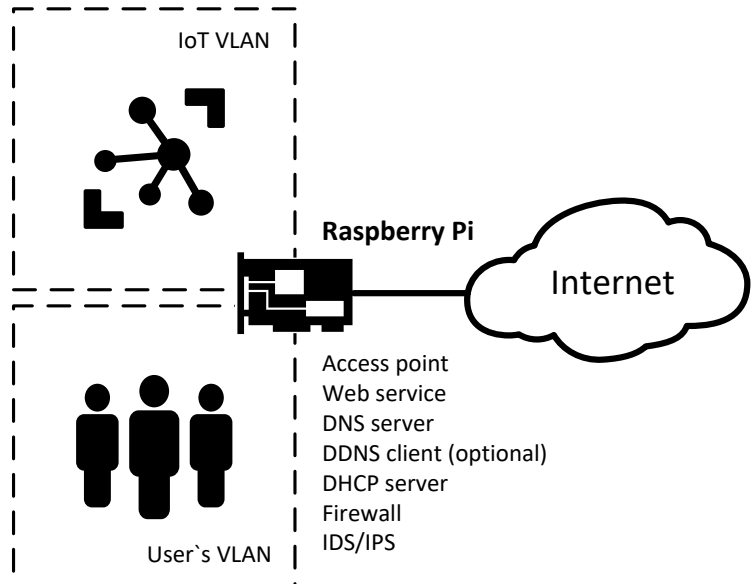

Fig. 4. Virtual networks on one device

In this scenario on Raspberry Pi be should install at least these services: web service (e. g., Apache); DNS server (Bind); DDNS client (optional); access point (HostAPd); DHCP server; firewall (iptables); IDS/IPS (Snort).

Also for balancing load and make another layer for security, it can be expand and add a router to another scenario (Fig. 5).

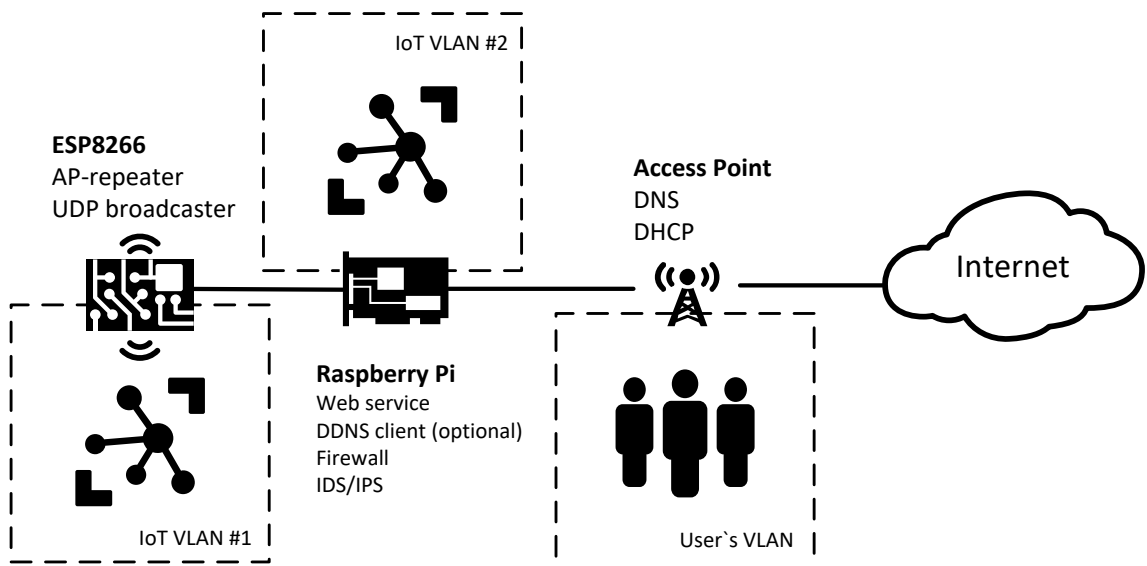

Fig. 5. Virtual networks on different devices

In this scenario on RaspberryPi be should install at least these services: web service (Apache); access point (HostAPd); firewall (iptables); IDS/IPS (Snort). On AP we need install: DHCP and DNS.

And in this scenario we don't need DDNS and also ESP8266 working as Access-point Repeater and it resend all UDP packets that received from other IoT devices who are connecting to that. In this way, Raspberry Pi can work as brain of this network and some IoT devices are connecting to Raspberry Pi directly and some others they are connecting to ESP8266, and ESP8266 repeating the Raspberry Pi wireless network and also it send all UDP Data's to Raspberry Pi then directly or in-directly all IoT devices are connecting to Raspberry Pi and it can analyze all of these data's and show results on Dashboard.